\documentclass[]{spie}  
\usepackage[]{graphicx}

\title{Thick CZT Detectors for Space-Borne X-ray Astronomy} 
\author{
H. Krawczynski\supit{a}, 
I. Jung\supit{a}, 
J. Perkins\supit{a}, 
A. Burger\supit{b}, 
M. Groza\supit{b}
\skiplinehalf
\supit{a}Washington University in St.\ Louis,  Department of Physics,\\ 
1 Brookings Dr.,  CB 1105, St. Louis, MO 63130; 
\skiplinehalf
\supit{b}Fisk University, Department of Physics,\\
1000 Seventeenth Ave. North, Nashville, TN 37208}
\authorinfo{Further author information: Send correspondence to H.K.: 
E-mail: krawcz@wuphys.wustl.edu, Telephone: 314 935 8553}
\begin{document} 
\maketitle 
\begin{abstract}
Cadmium Zinc Telluride (CZT) detectors are having a major impact on the field
of hard X-ray astronomy. Without the need for cryogenic cooling they achieve
good spatial and energy resolutions over the broad energy range from 10 keV 
to $\sim$600 keV. In this paper, we briefly review the historical development 
of detectors used in X-ray astronomy.
Subsequently, we present an evaluation of CZT detectors from the company
Imarad. The standard 2$\times$2$\times$0.5~cm detectors, contacted 
with 8$\times$8 In pixels and an In cathode, exhibit FWHM energy resolutions 
of 7 keV  at 59 keV, and 10 keV at 662 keV. A direct measurement of the 
662~keV photopeak efficiency gives 67\%. 
We have started a detailed study of the performance of Imarad detectors 
depending on surface preparation, contact materials, contact deposition, 
post-deposition detector annealing, and detector passivation techniques. 
We present first results from contacting detectors with Cr, Ag, Au, and Pt.
\end{abstract}
\section{Introduction -- detectors in X-ray astronomy}
X-ray astronomy has come a long way since the first detection of X-ray emission from 
an extra-solar object in 1962\cite{Giac:62}. 
In the meantime, X-ray astronomy has emerged as one of the drivers of modern astronomy, 
as evident from the award of the Nobel prize in physics to X-ray pioneer Riccardo 
Giacconi in 2002, and the existence of two major X-ray observatories, 
{\it Chandra} and {\it XMM-Newton}, orbiting Earth since 1999.

%
%
The most important results of X-ray astronomy have so far been obtained from observations 
in the ``soft'' X-ray band with photon energies between 0.1 keV and several keV. 
To mention several important milestones, the {\it Einstein} observatory\cite{Giac:79} 
achieved in 1978 a first breakthrough in the 0.2 keV to 4 keV band by 
combining a grazing incidence mirror with position sensitive detectors. 
The observatory carried a ``High Resolution Imager'' (HRI), consisting of two cascaded Micro 
Channel Plates (MCP) read out by a crossed-grid charge detector. 
Another main instrument was the Imaging Proportional Counter (IPC), a 4 cm deep, 
800 Torr Ar-Xe-CO$_2$ proportional counter which achieved a modest energy resolution 
of between 50\% and 100\%. 
The grazing incidence mirror and the HRI achieved an effective 
detection area of 10 cm$^2$. The mirror and the IPC had a detection area of 100 cm$^2$.
The {\it Roentgen Satellite (ROSAT)} was launched in 1990 and used similar detectors as 
the {\it Einstein} observatory. Its HRI and IPC cameras collected
respectively 8 and 2.4 times more photons per unit time than their {\it Einstein} 
counterparts\cite{True:83}.
The {\it Chandra} and {\it XMM-Newton} satellites are equipped with 
powerful X-ray grazing incidence mirrors and state-of-the-art photon detectors. 
The {\it Chandra} satellite carries the {\it Advanced CCD imaging spectrometer (ACIS)}\cite{Garm:03} 
made of Si-CCDs with energy resolutions of between 2\% and 10\% 
over the energy range from 0.1 keV to 10 keV. 
The {\it XMM-Newton} satellite carries two conventional Metal Oxide Semiconductor (MOS) cameras, 
and a ``PN'' camera\cite{Stru:01}. The fully depleted $pn$ CCDs (280 micron deep 
depletion region) can detect X-rays with energies up to 15 keV. 
The {\it Chandra} X-ray mirror and ACIS camera achieve a 1 keV detection area of 340 cm$^2$ and 
an unique angular resolution of  0.5''. The {\it XMM-Newton} mirrors and CCD cameras have a 
combined 1 keV detection area of 2,100 cm$^2$ and an angular resolution of 15''.

%
%
Compared to the soft X-ray band, the hard X-ray band ($\sim$ 20-200 keV) has progressed more slowly.
Hard X-rays are more difficult to focus, scintillation detectors had only modest
spatial and energy resolutions, and Ge detectors required resource intensive cryogenic cooling. 
The {\it High Energy Astrophysics Observatory-1 (HEAO-1)} was equipped with 
a 1.4$^\circ$ $\times$ 20$^\circ$ fan collimator and 250~cm$^2$ of NaI scintillator, 
and scanned the hard X-ray sky in the years 1977-1979 \cite{Matt:78}.
More recently, the {\it Rossi X-Ray Timing Explorer (RXTE)}\cite{Brad:90} launched in 1995 
and {\it Satellite per Astronomia X, ``Beppo'' (BeppoSAX)}\cite{Piro:95} launched in
1996 used essentially the same detector technology with 1$^\circ$ pencil beam collimators.
The wide bandgap semiconductors Cadmium Telluride (CdTe) and Cadmium Zinc Telluride (CZT)
are now boosting the field of hard X-ray astronomy. The {\it International Gamma-Ray Astrophysics Laboratory 
(INTEGRAL)}\cite{Uber:03} launched in 2002 carries the {\it Imager on Board the Integral Satellite (IBIS)} 
that uses CdTe detectors with a detector area of $\sim$2,600 cm$^2$.
The {\it SWIFT} satellite\cite{Gehr:00}, to be launched in October 2004, carries the 
{\it Burst Alert Telescope (BAT)} with a CZT detection area of $\sim$5,240 cm$^2$.
The {\it IBIS} and the {\it BAT} use the approach of coded mask imaging.
Embedded in a collimator assembly, the detectors see the sky through a 
patterned shadow mask. 
The detectors image the sum of all shadows cast by the X-ray sources in the field of view.  
A deconvolution algorithm is used to derive the X-ray surface brightness distribution
from the detected image. 
In this approach, the effective detection area is roughly half of the 
active detector area (the other half of the detector area is covered by the 
coded mask), and is not given by the collection area of any focusing optics.

The congressionally approved Beyond Einstein Program\cite{BeyondEinstein} of NASA's
``Structure and Evolution of the Universe'' theme plans for two major X-ray missions. 
In both missions, CZT plays an important role.
The Constellation-X observatory \cite{conX} will perform high throughput X-ray spectroscopy
with high energy and angular resolution over the energy range from 0.25 keV to 10 keV employing
X-ray gratings with Si-CCDs, as well as non-dispersive microcalorimeters.
The design includes a hard X-ray telescope that uses CZT detectors to cover the energy range from
6~keV to 40 keV or higher. 
Constellation-X is currently receiving NASA funds for Pre-Phase A development, 
the ``advanced concepts'' phase. Given the current budget plan, the mission will be 
launched in 2016\cite{Sala:04}.
A second recommended mission is the ``black hole finder probe'', a large field of view, 
hard X-ray telescope. The {\it Energetic X-ray Imaging Survey Telescope (EXIST)}
design\cite{Grin:03} proposes an assembly of coded mask imagers.
Hard X-rays are detected with $\sim$0.5~cm thick CZT detectors 
with a total detector area of $\sim$80,000 cm$^2$. 
{\it EXIST} has been selected for a 2-year NASA concept study, along with 
the {\it Coded Aperture Survey Telescope for Energetic Radiation 
(CASTER)}\cite{McCo}, a competitor design which employs the new 
scintillators LaBr$_3$ or LaCl$_3$ as primary detectors. 
There is no definitive time schedule for the black hole finder probe, 
as its funding is presently beyond the budgetary horizon of 2009\cite{Sala:04}. 

In this paper, we emphasize thick CZT detectors and their application in {\it EXIST}-type 
hard X-ray telescopes. In Sect.\ \ref{requirements} we will discuss key requirements for
CZT detectors in this application. In Sect.\ \ref{standard} we scrutinize the performance 
of In contacted standard detectors from Imarad.
Furthermore, we present detailed detector simulations and compare simulated data 
with experimentally measured data. In Sect.~\ref{contacts} we give first results 
of contacting the detectors with Cr, Ag, Au, and Pt.
Finally, in Sect.~\ref{summary}, we summarize our findings and describe work in progress.
\section{Thick CZT for hard X-ray telescopes}
\label{requirements}
\subsection{CZT substrates}
Presently, major vendors of CZT substrates are eV Products\cite{eV}, Imarad\cite{Imarad}, and Bicron\cite{Bicr}.
eV Products and Bicron use the High-Pressure Bridgman (HPB) process  to grow the substrates, 
while Imarad uses the modified High-Pressure Bridgman (mHPB) process.
While HPB and mHPB CZT achieve similar 
performances at high energies, mHPB detectors give poorer results at low energies. 
The latter can in part be explained  by the higher resistivity of 
HPB crystals ($\sim$2$\cdot 10^{11}$ ohm-cm) compared to
mHPB crystals ($\sim$$10^{10}$ ohm-cm) that results in lower leakage currents.
An alternative to CZT might be CdTe\cite{Taka:02}. 
An advantage of CdTe is that large uniform single crystals can be grown.
Although CdTe detectors require cooling to temperatures of -20$^\circ$C 
to achieve good performances, such temperatures can easily be maintained in space.
\subsection{Requirements for thick CZT detector in space-borne X-ray telescopes}
In the following, we discuss the requirements for 0.5~cm thick CZT detectors used for the detection of 
10 keV to 600 keV photons in space-borne X-rays telescopes. \\[1.5ex]
%
%
{\it Photopeak efficiency:} 
A crucial detector parameter is the photopeak efficiency $\epsilon$.
The detector area required to achieve a certain flux sensitivity
scales with $1/\epsilon$; the mass and cost of the full satellite can depend even more
strongly on $\epsilon$. Even in a fully depleted detector and after full Depth of Interaction (DOI) 
correction, not all photo-effect events show up in the measured energy spectra. 
The main culprit is most probably ``dead detector regions''.
Owing to non-negligible surface conductivity, not all electric field lines end at the 
anode pixels: some end in-between pixels\cite{Bolotnikov01}.
Electron clouds drifting to the area between pixels induce less charge on the anode contacts
than clouds drifting to the anode contacts themselves.
While large pixels minimize the dead volume, they usually do not give ideal energy resolutions. 
The volume of the dead regions can be reduced by passivating the detectors after contact 
deposition. Another approach is to use negatively biased steering contacts located between 
the anode contacts to steer the electrons toward the anode contacts.
\\[1.5ex]
%
%
{\it Energy and spatial resolutions:} 
The best reported energy resolutions of 0.5 mm thick CZT detectors are about 3 keV at 59 keV
and 7 keV at 662 keV\cite{Macr:02,Nari:99,Perk:03}. 
The {\it EXIST} design foresees pixellated detectors with a pixel pitch of about 1.25~mm to achieve 
an angular resolution of between 2' and 5'.
Pixellated detectors are preferred, as they exhibit lower lead capacitance and lower leakage 
currents than strip detectors. 
DOI resolution of 1\% or better is required to assure that the anode signals can
be corrected for any existing DOI dependence.
An important fact to notice is that energy resolutions  below $\sim$100 keV 
are typically dominated by electronic noise (produced by the detector itself) and readout noise. 
The energy resolutions above $\sim$ 100 keV are dominated by crystal
inhomogeneities and the dependence of the induced charge on the location 
of the photon absorption.
Thus, a detector performing well at low energies does not necessarily perform well at 
high energies and vice versa. 
While present CZT technology gives high-energy resolutions 
that are sufficient for most astrophysical applications, a substantial improvement 
of the low-energy resolution would be very welcomed.\\[1.5ex]
{\it Temperature range and mechanical ruggedness:}
In space, CZT detectors can conveniently be operated at temperatures between -50$^\circ$C and 0$^\circ$C. 
Most CZT detectors show better performance at such low temperatures than at room temperatures. 
However, some Imarad mHPB detectors equipped with blocking contacts basically stopped working at 
temperatures of about -10$^\circ$C\cite{Nari:00}.
The detectors should also be rugged enough to withstand vibrations during the launch of the spacecraft. 
\\[1.5ex]
\begin{figure}[tb]
\vspace*{-0.4cm}
\begin{center}
{\includegraphics[width=12.6cm]{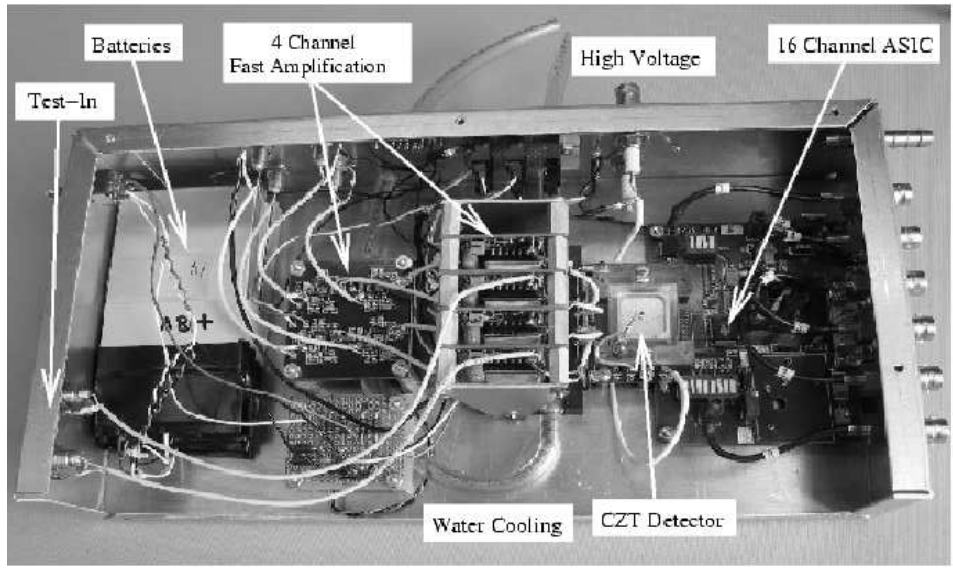}}
\end{center}
\vspace*{-0.4cm}
\caption{\label{mounting} Test set-up for evaluation of the CZT detector performance (Washington University). 
The hybrid read-out system consists of a low-noise ASIC readout for 16 pixels and a fast 
(100 MHz analog bandwidth) readout for 3 pixels and the cathode; pixels which are not read out
are held at ground potential.}
\vspace*{-0.4cm}
\end{figure}
%
%
{\it Radiation hardness:}
Space-borne detectors suffer year long exposures to trapped charged particles, 
cosmic rays, and neutrons that damage the detectors and make them radioactive. 
Several groups have studied the radiation damage of CZT detectors produced by high 
energy protons\cite{Varn:96,Wong:96,Hull:97,Fran:99,Slav:00,Mura:03}.
A 200 MeV proton dose of  $\sim$10$^9$ cm$^{-2}$ decreases the electron 
$\mu-\tau$ product by 20\% to 40\% and reduces the charge yield by 
10\% to 20\%\cite{Wong:96,Slav:00}.
Wong et al.\ (1996)\cite{Wong:96} reported that annealing the detectors 
to about 100$^\circ$C can be used to undo the damage.
Fraboni et al.\ (2003)\cite{Frab:03} irradiated CZT detectors with large doses of
$\gamma$-rays, electrons (9~MeV), thermal ($\sim$1 eV) and fast ($\sim$1 MeV) neutrons, 
and protons (2 MeV). Using photo induced current transient spectroscopy they 
find that the degradation of the detector performance (increase in leakage current, 
loss of energy resolution,  and shift of the photopeak to lower energies) 
is linked to the appearance of very specific deep energy levels.
\\[1.5ex]
{\it Activation:} Substrate activation is a serious concern for thick detectors.
A number of background measurements have been made in several 
balloon flights\cite{Pars:96,Slav:00,Jenk:03}. 
Slavis et al.\ (2000)\cite{Slav:00} monitored the background during
a 21~hrs long 104,000 ft high balloon flight and used the data to 
derive upper limits on the activation of a CZT detector. 
Murakami et al.\ (2003)\cite{Mura:03} activated a CdTe detector with 155 MeV protons and 
subsequently identified a large number of radio-active elements with a 
Ge detector and the CdTe detector itself.
Although several Monte Carlo studies of CZT activation on different time scales
have been published\cite{Arms:99a,Arms:99,Perf:01,Blos:02}, there are no
long-term activation data to test the simulations.
The {\it INTEGRAL} and {\it SWIFT} missions will provide data sets to 
improve on this situation.
\section{Performance of standard Imarad detectors}
\label{standard}
In this section, we discuss the performance of standard Imarad
detectors. The 2$\times$2$\times$0.5~cm detectors are contacted with
8 $\times$ 8 In pixels on the anode side and a monolithic In cathode.
The pixels have a diameter of 1.6~mm and the pixel pitch is 2.4 mm.
The detectors have a conducting band around their perimeter.
Even when left floating, the band improves the photopeak efficiency 
and energy resolution of the outer pixels. 
We concentrate on the performance of central pixels.
Previous measurements of Imarad detectors have been described by  
Narita et al.\ (1999, 2000)\cite{Nari:99,Nari:00}, Li et al.\ (2001)\cite{Li},
Nemirovski et al.\ (2001)\cite{Nemi:01}, 
Hong et al.\ (2002)\cite{Hong:02}, 
and Perkins et al.\ (2003)\cite{Perk:03}. 
\subsection{Mounting and electronic readout}
We mount the detectors using gold plated pogo-pin contacts on both, the anode
and the cathode sides. 
The cathode is negatively biased, and the anode pixels are held at ground.
We use a hybrid electronic readout (see Fig.\ \ref{mounting}). 
Four channels (3 anode pixels plus the cathode) are AC coupled and are amplified 
by a fast Amptek 250 amplifier followed by a second amplifier stage. 
The amplified signals are digitized by a 500 MHz oscilloscope 
and transferred to a PC via Ethernet. 
The time resolved readout enables us to measure the drift time 
of electrons through the detector to a resolution of 10 ns.
We use a 16 channel ASIC from eV products\cite{eV} to measure pulse height 
information for 16 additional pixels. The ASIC gives amplified and shaped 
signals which we digitize with custom designed VME boards.
The FWHM noise of both readout chains lies between 5 keV and 10 keV.  
If not stated otherwise, we quote all energy resolutions after quadratic
subtraction of the readout noise determined at zero detector bias.
\subsection{Individual pulses}
\begin{figure}
  \centering
  \begin{minipage}{3.5in}
  \includegraphics[width=3.4in]{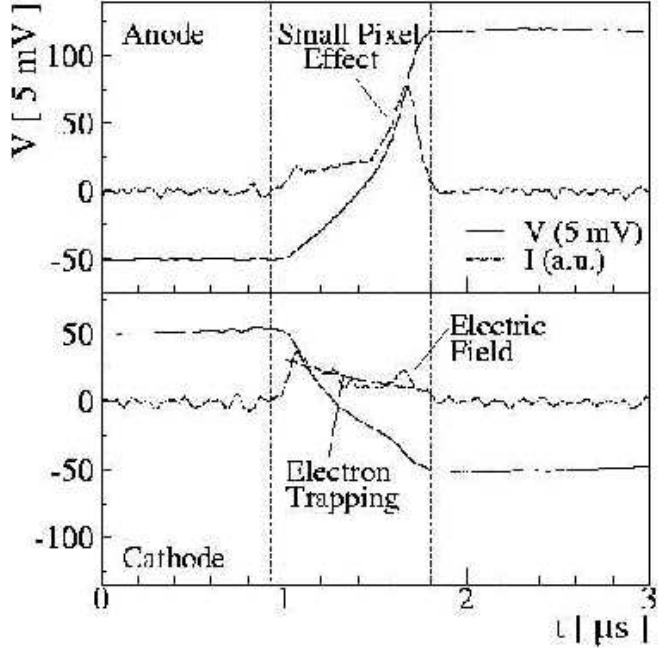}
  \end{minipage}
  \begin{minipage}{2.8in}
  \caption{\label{p1}
  The panels show an anode (top) and cathode (bottom) pulse of a 662 keV photon measured
  with a standard Imarad detector. In both panels, the solid line is proportional to the 
  charge induced on the contact while the dashed line is the time derivative of that charge. 
  The vertical lines mark the beginning and end of the pulse.  
  The anode current peaks toward the end of the pulse owing to the small pixel effect.
  The cathode current decreases with time, probably owing to electron trapping.
  The detector was biased at -400 V.}
  \end{minipage}
\end{figure}
Figure \ref{p1} shows pulses induced by a 662 keV (Cs$^{137}$) photon at a detector bias of -400~V. 
The time derivative of the charge sensitive preamplifier shows the induced current. 
The current rises slowly as the electrons drift toward the anode pixel.
At the end of the 0.5~$\mu$sec pulse, a pronounced current peak can be 
recognized. The peak is caused by the rapidly decreasing weighting
potential near the pixels owing to the small pixel effect\cite{Barret,Luke}.
Electron trapping is seen in an exponential decrease of the cathode current.

\begin{figure}
  \centering
  \hspace*{-0.9cm}
  \begin{minipage}{3.4in}
  \includegraphics[width=3.8in]{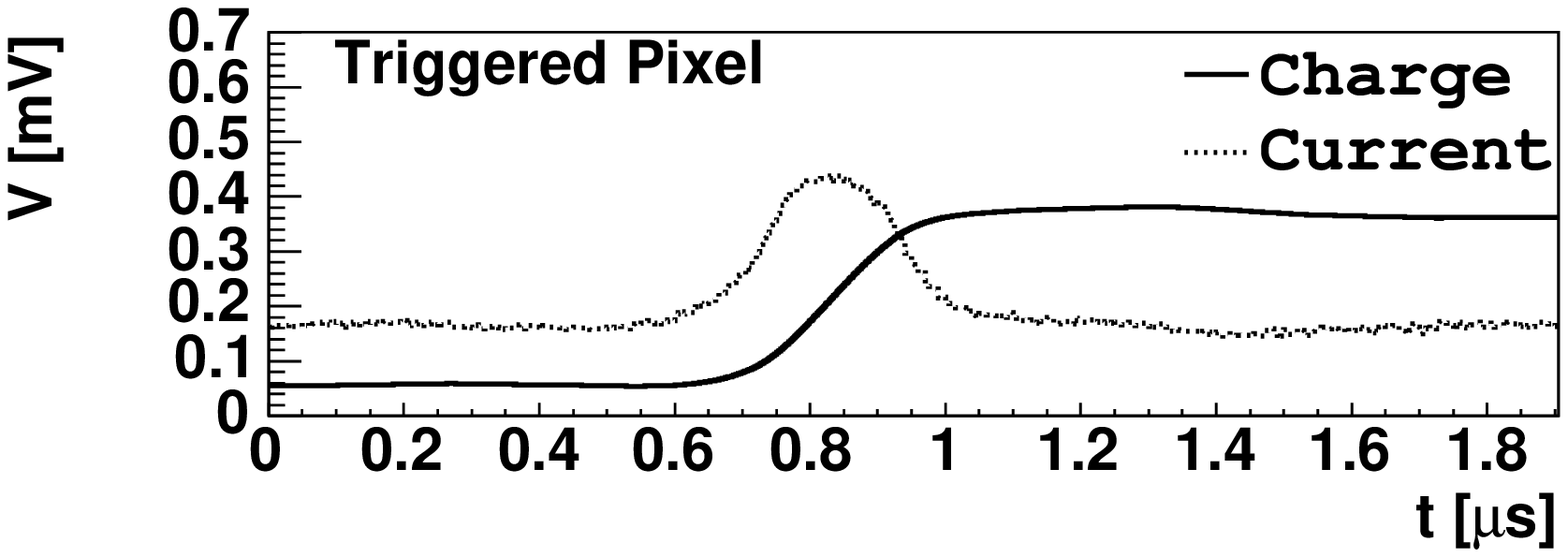}
  \end{minipage}
  \begin{minipage}{3.4in}
  \includegraphics[width=3.8in]{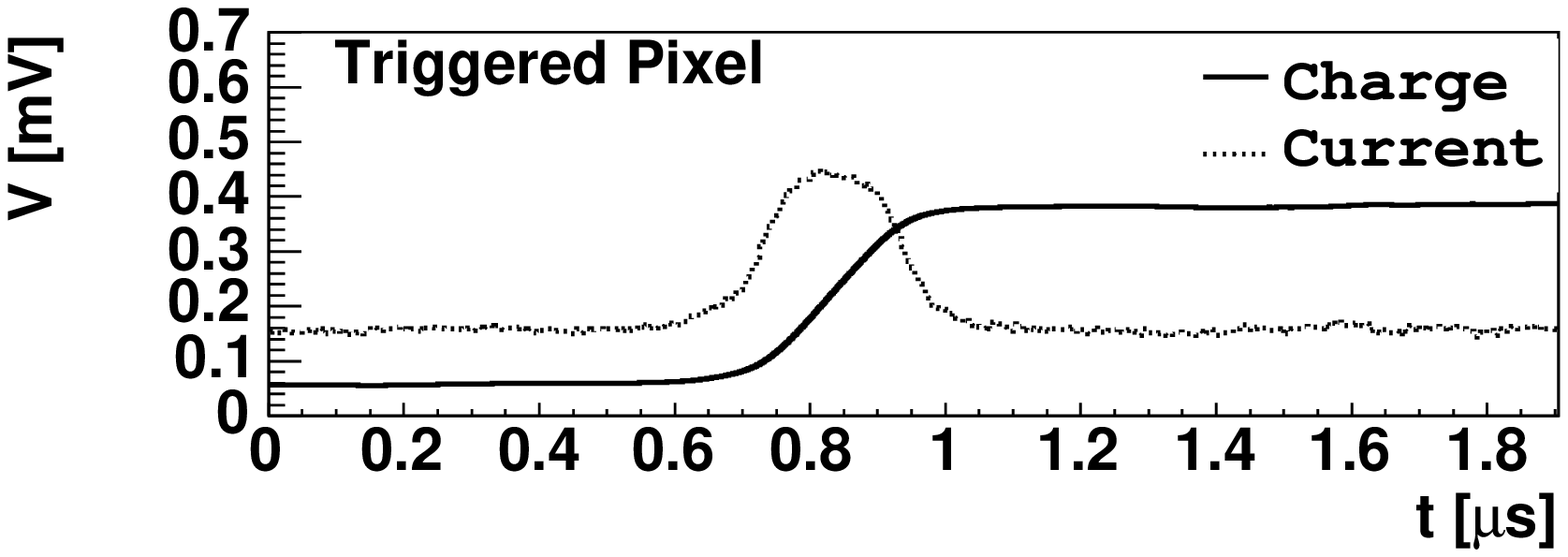}
  \end{minipage}\\[-.2cm]
  \hspace*{-0.9cm}
  \begin{minipage}{3.4in}
  \includegraphics[width=3.8in]{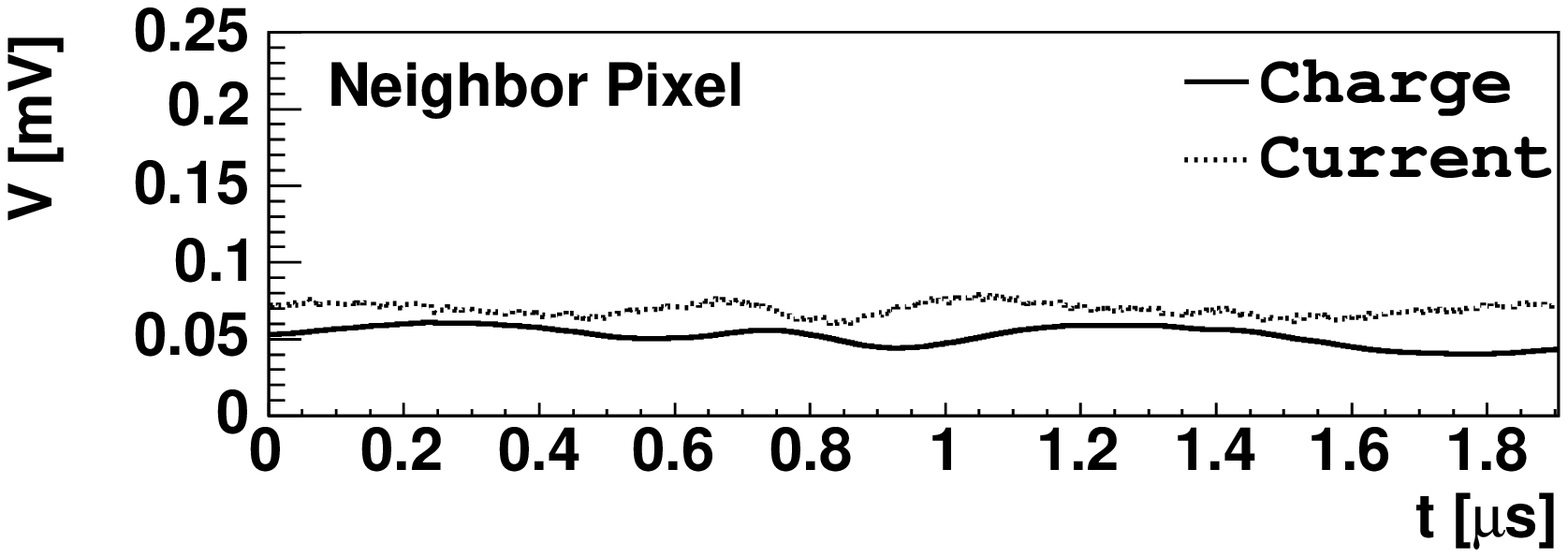}
  \end{minipage}
  \begin{minipage}{3.4in}
  \includegraphics[width=3.8in]{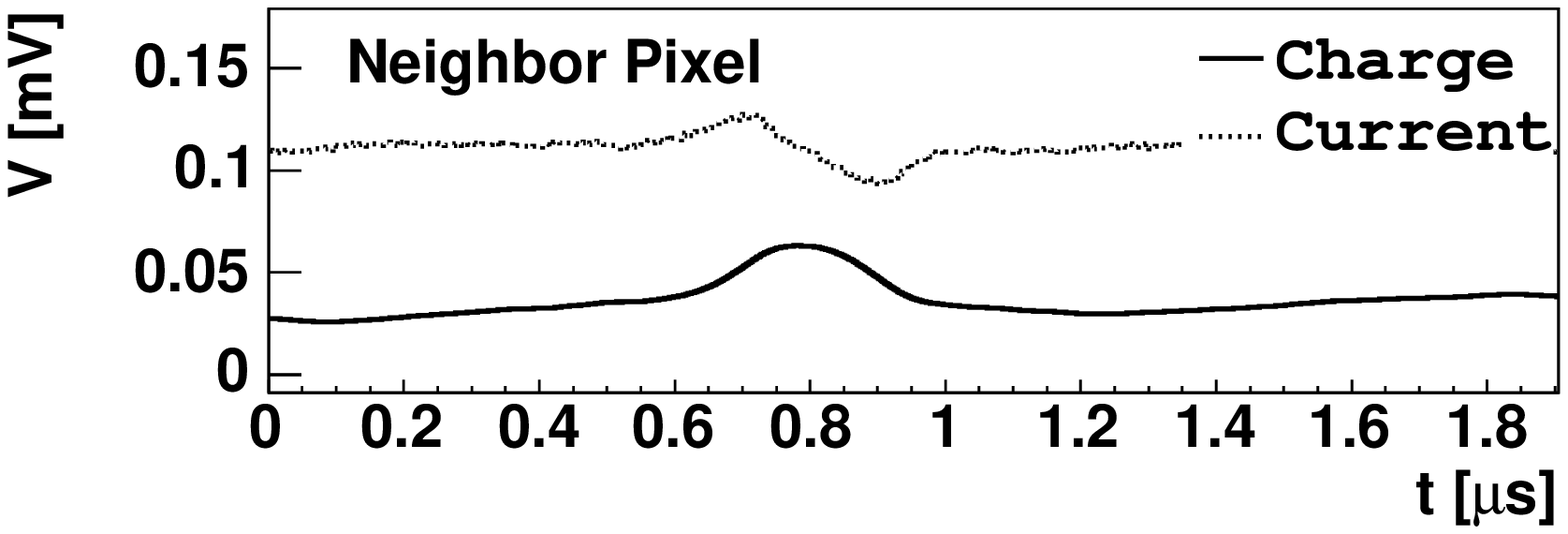}
  \end{minipage}\\[-.2cm]
  \hspace*{-0.9cm}
  \begin{minipage}{3.4in}
  \includegraphics[width=3.8in]{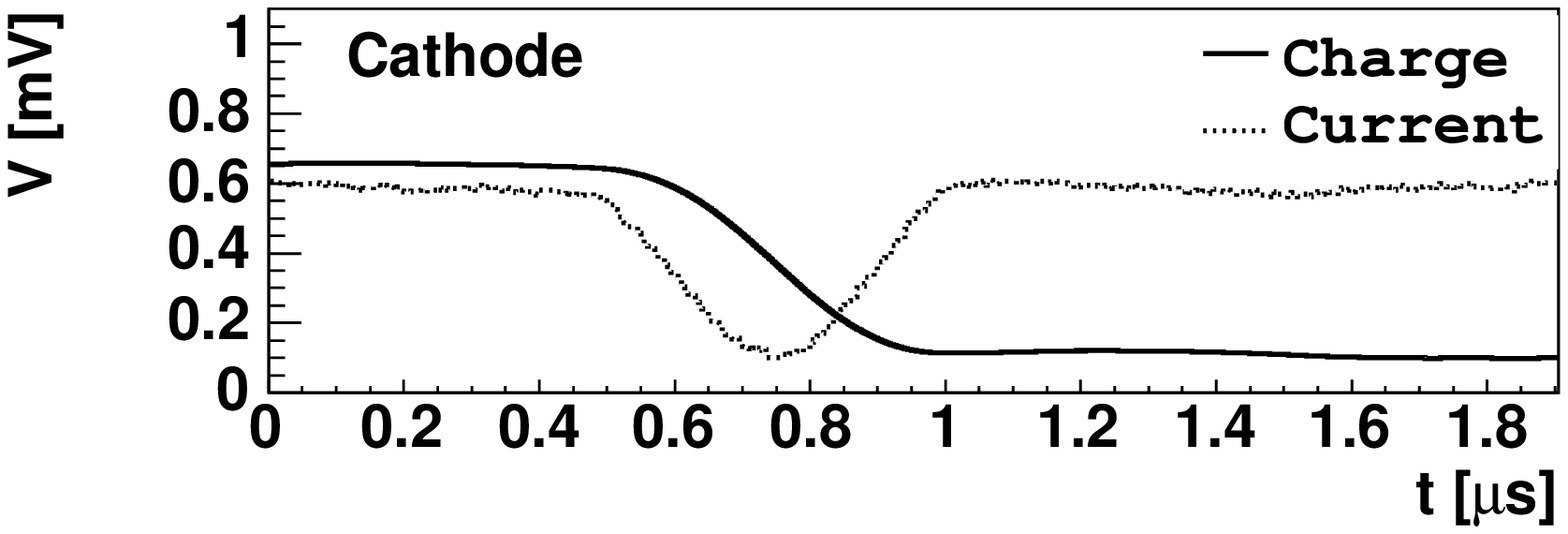}
  \end{minipage}
  \begin{minipage}{3.4in}
  \includegraphics[width=3.8in]{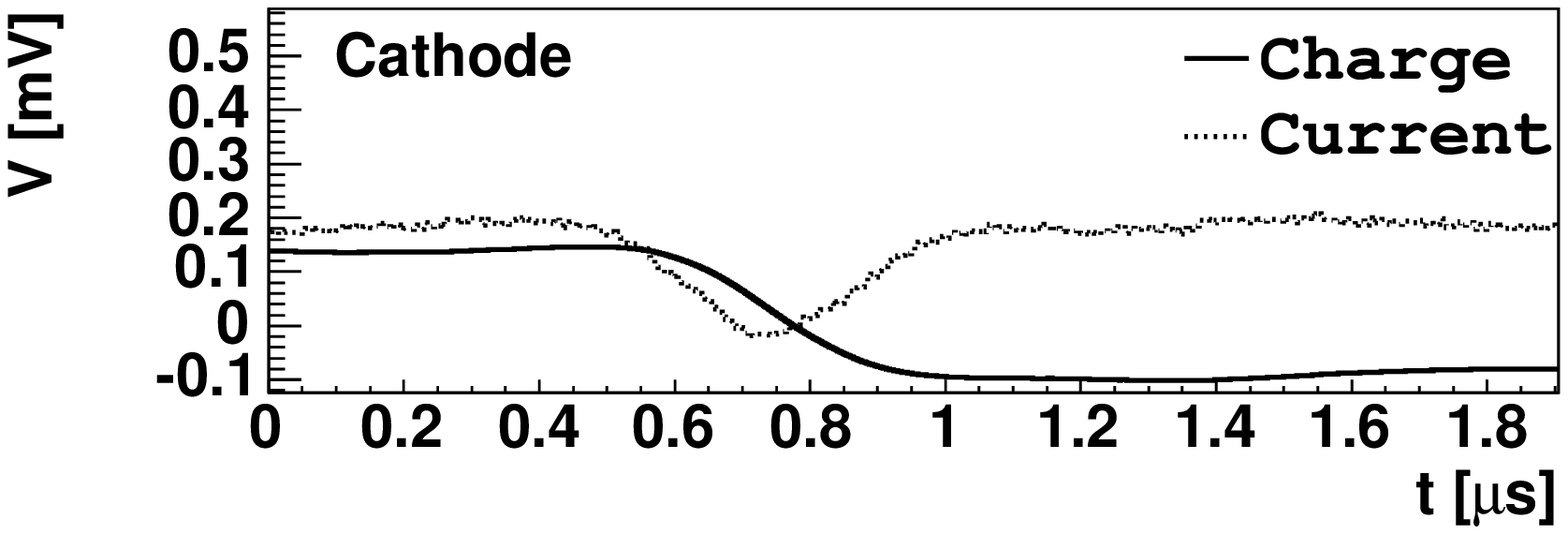}
  \end{minipage}
  \caption{  \label{p2}
  Two 662 keV events measured with a standard Imarad detector.
  The left side shows an event with no signal in any neighboring pixel and the right side
  shows an event with a relatively large signal in a neighboring pixel.
  The solid lines are proportional to induced charge, and the dotted lines 
  are proportional to the induced current.
  The neighboring pixel on the right side shows a bipolar pulse, as
  expected when the electrons do not drift to the pixel itself.
  The detector was biased at -1000 V.}
\end{figure}

Figure \ref{p2} shows two events induced by 662 keV photons at a detector bias of -1000~V.
The event on the left side has no signal in any neighboring pixel, while the event on the right side 
has a neighboring pixel with a relatively large signal.
The triggered pixel shows a large unipolar pulse for both events.
In the second event, the adjacent pixel shows a bipolar pulse, 
as expected when the electrons do not drift to that pixel.
We are currently studying the possibility to use the information encoded in 
bipolar pulses to improve the photopeak efficiency and energy resolution of the 
detectors.
\subsection{Energy spectra}
\begin{figure}
\begin{center}
\begin{minipage}{3.2in}
 \includegraphics[width=2.6in]{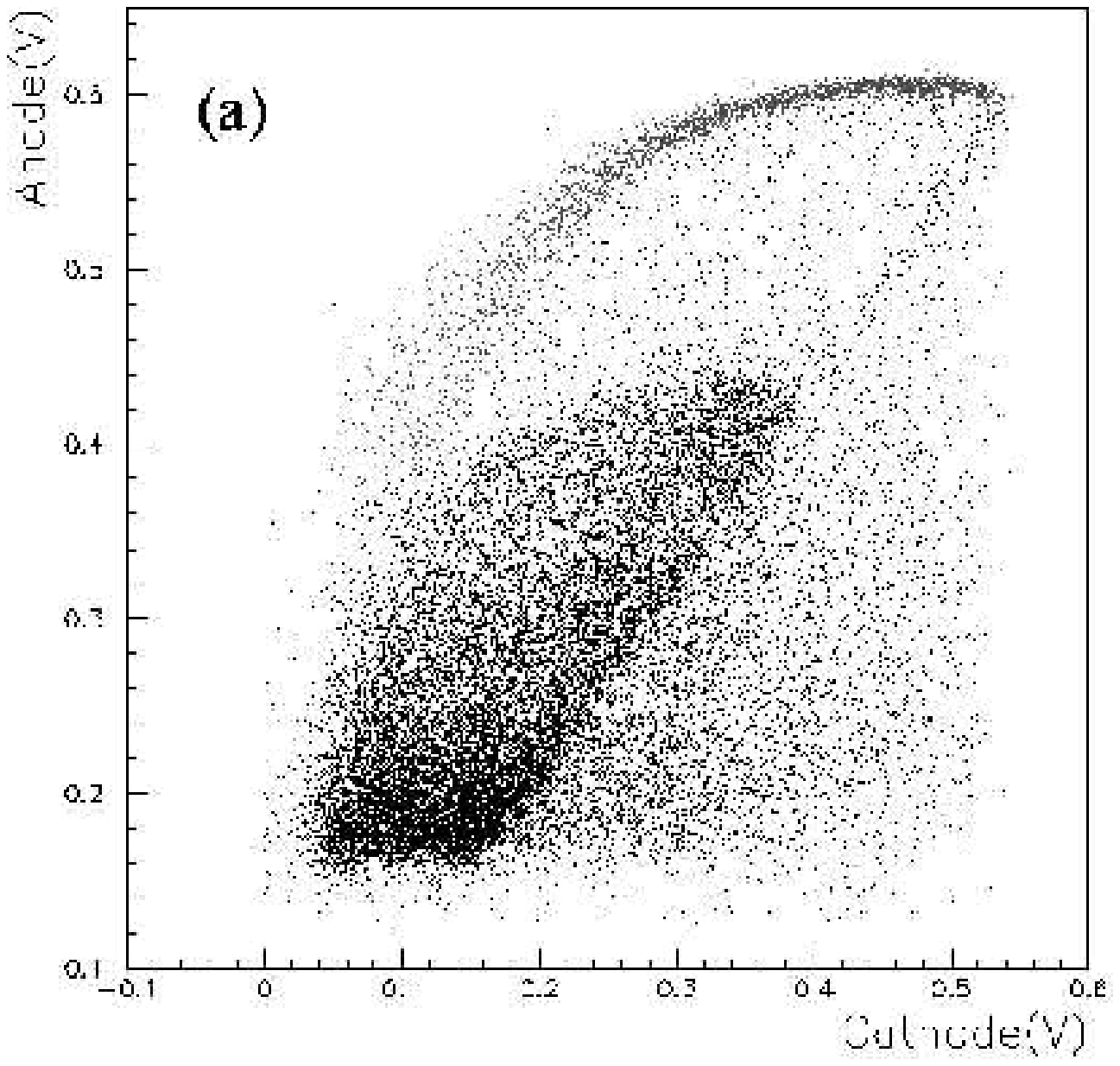}
\end{minipage}
\begin{minipage}{3.2in}
 \includegraphics[width=2.6in]{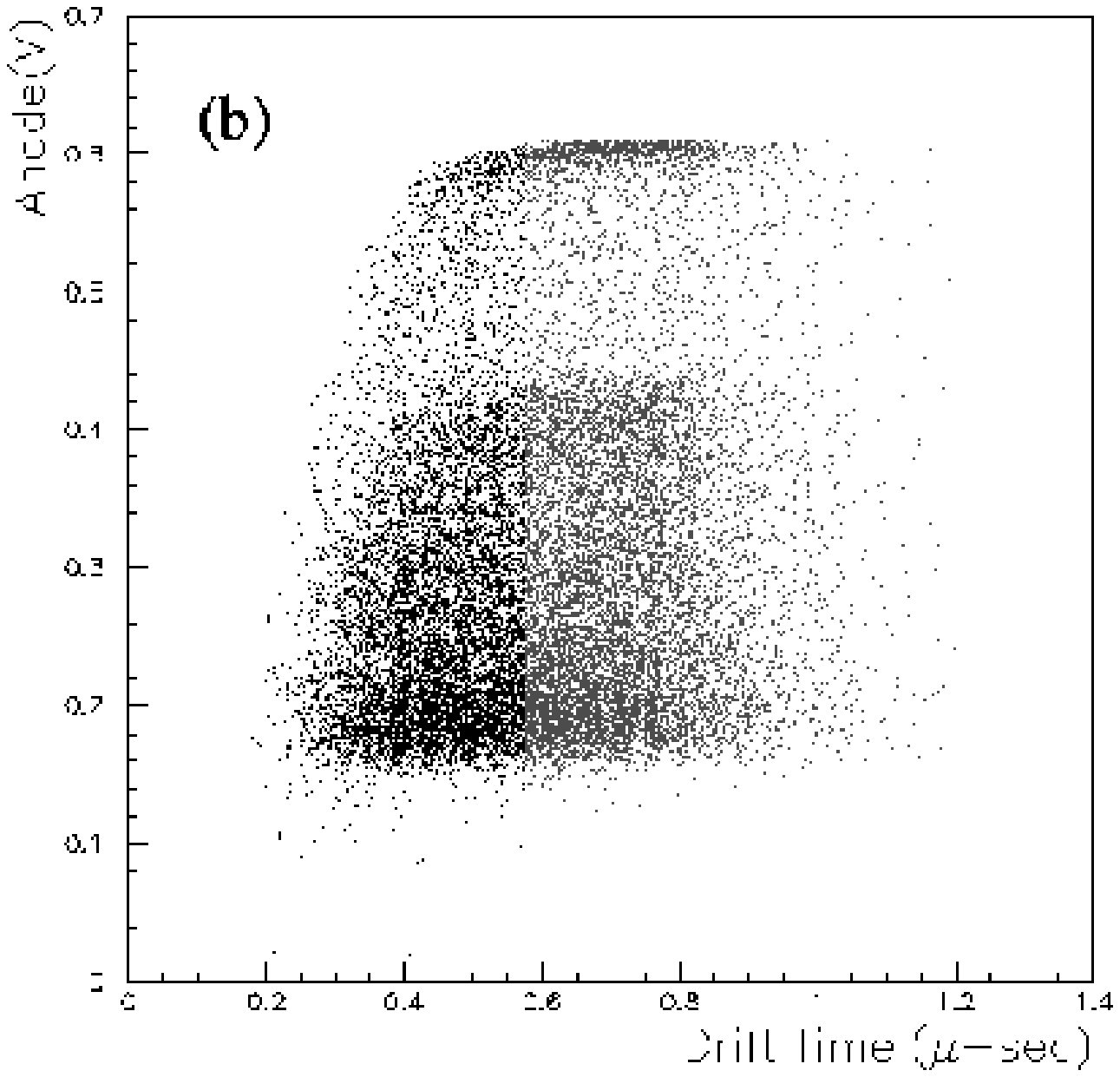}
\end{minipage}
\begin{minipage}{3.2in}
 \includegraphics[width=2.6in]{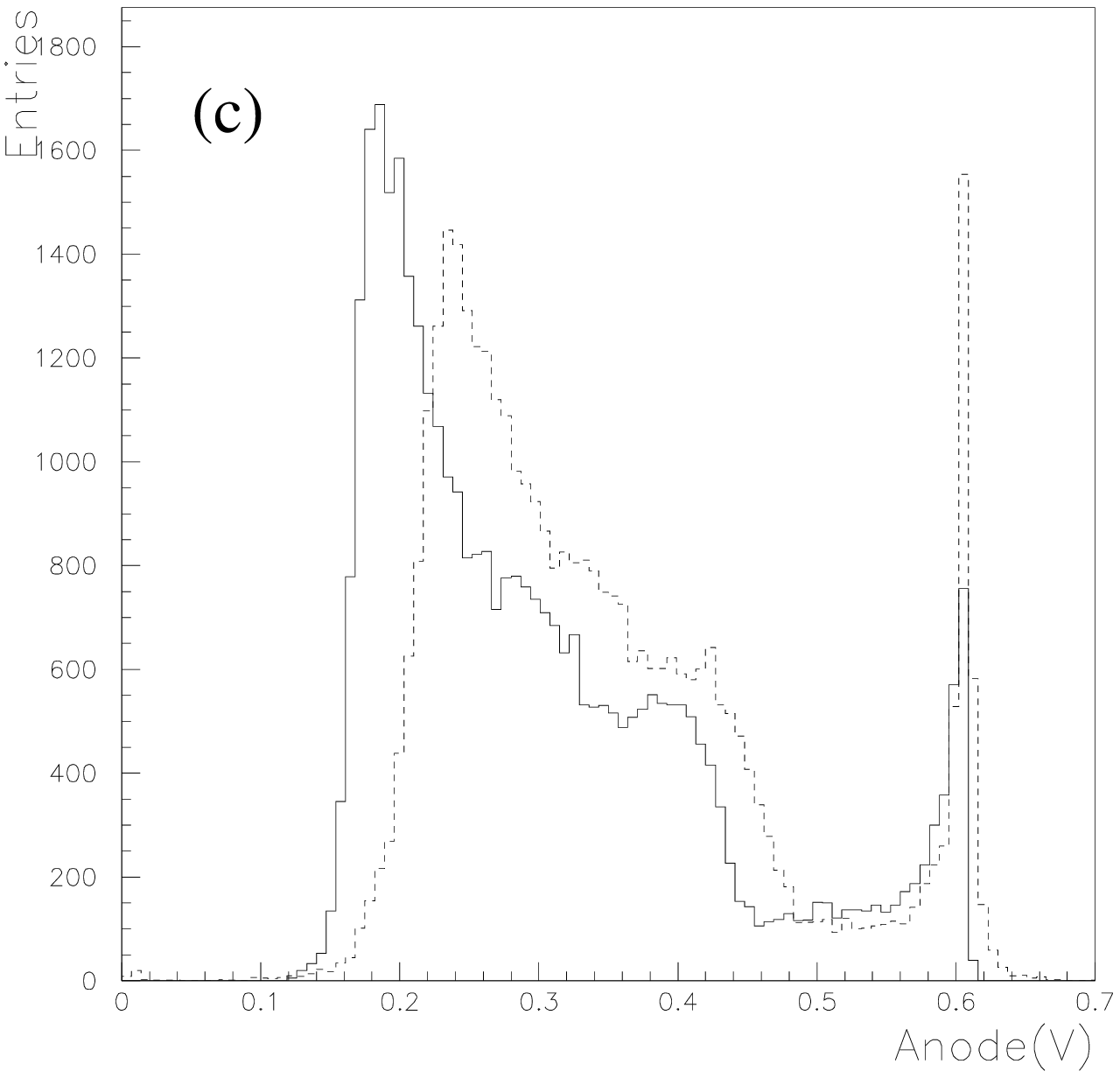}
\end{minipage}
\begin{minipage}{3.2in}
 \includegraphics[width=2.6in]{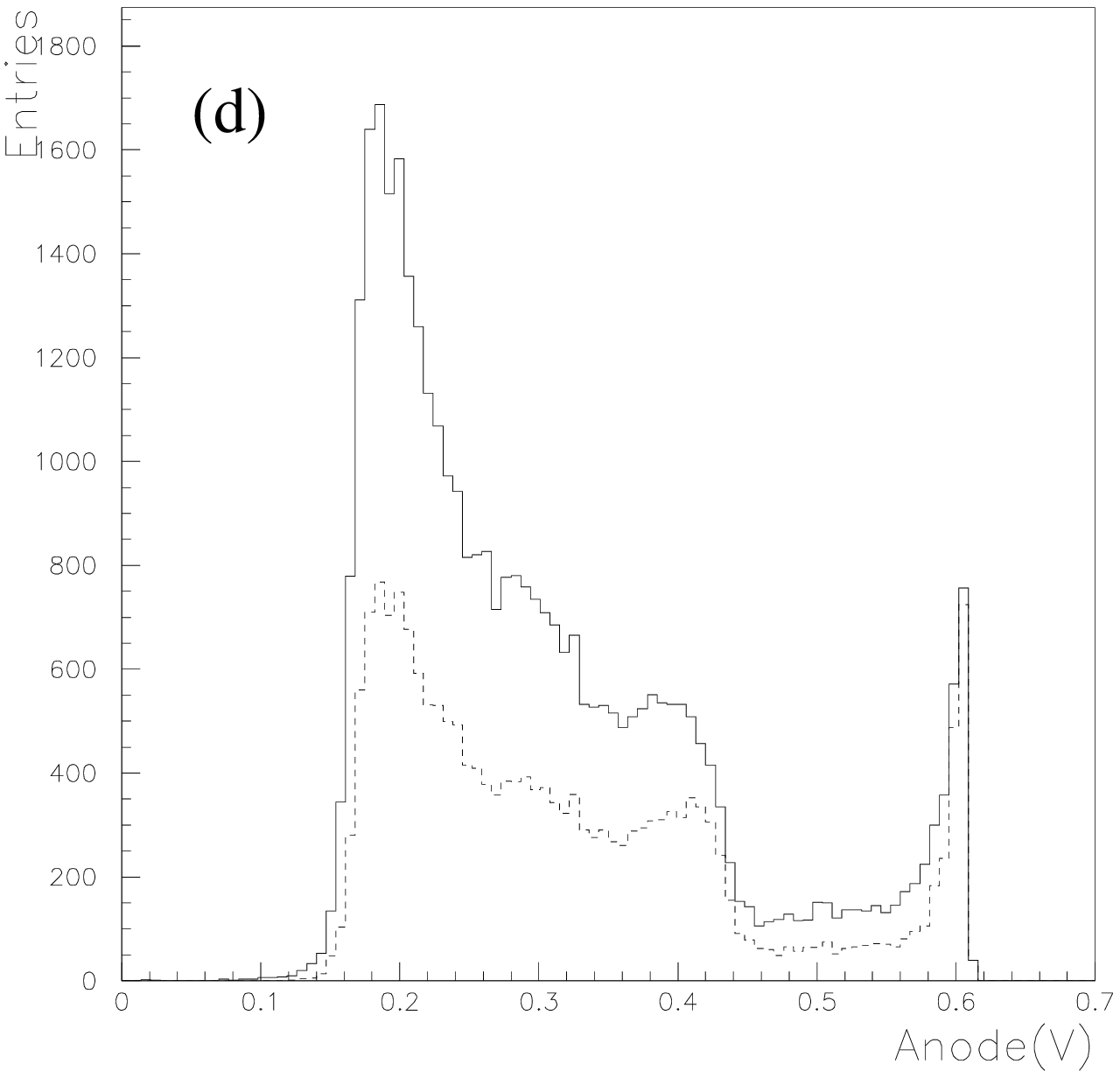}
\vspace*{-2ex}
\end{minipage}
\end{center}
  \caption{  \label{spectra}
  Panel a shows the correlation of the anode amplitude with the  
  cathode amplitude measured with a standard Imarad detector (662 keV). 
  Panel b shows the correlation of the anode amplitude with the duration
  of the cathode pulse (electron drift time). 
  Panels c and d show energy spectra before (solid line) and after 
  (dashed line) DOI correction with Method 1 and 2, respectively.
  The detectors were biased at $\sim$-1000 V.\vspace*{-2ex}
}
\end{figure}

Owing to hole trapping, the induced anode signals depend on the DOI of the photon. 
The effect is mitigated by the small pixel design of the detector. 
While the effect is negligible at energies below $\sim$100 keV (when all photons interact close to the cathode), 
it becomes increasingly important at higher energies when the photons start penetrating the 
substrate. We compared two different methods to correct the anode signals for their DOI dependence (Fig. \ref{spectra}). 
Method 1 uses the correlation of the anode and cathode amplitudes 
(panel a). Method 2 uses the correlation of the anode amplitude 
and the electron drift time as estimated from the length of the cathode pulse
(panel b). Method 1 improves the photopeak efficiency 
by 57\% and the FWHM energy resolution from 11 keV to 10 keV (panel c).
Method 2 does not work as well as Method 1. A cut in drift times $t_{\rm drift}\,>$ 0.57~$\mu$sec 
does improve the energy resolution from 11 keV to 10 keV, but
decreases the photopeak efficiency by 14\% (panel d).

At 59 keV, no DOI correction is needed as all photons interact near the cathode. 
We measured a 59 keV energy resolution of 7 keV.
Remarkably, the absolute resolutions at 59 keV and 662 keV are rather similar (7 keV and 10 keV, respectively).
The result indicates that the resolution is dominated by electronic noise (produced by the detector) 
rather than by detector inhomogeneities.
At -1000 V bias, we find typical leakage currents $I_{\rm d}$ between 
7 nA and 14 nA. The associated shot noise\cite{Nemi:98} is:
\begin{equation}
{\rm FWHM}_{\rm sh}\,=\,2.35\,\bar{\omega}\,
\frac{e}{\sqrt{\pi}}\,\sqrt{ \frac{I_{\rm d}\,\tau_{\rm s}}{q}}
\end{equation}
\begin{equation}
\,\,\,\,\,\,\,\,\,\,\,\approx\,4.1\,{\rm keV}\,
\left(\frac{I_{\rm d}}{\rm 10~nA}\,\frac{\tau_{\rm s}}{1\, \mu {\rm sec}}\right)^{1/2}
\end{equation}
where $\bar{\omega}\,\approx$ 4.5 eV is the mean energy required to create an
electron hole pair in CZT, $e\,=$ 2.718, $q$ is the electron charge in 
Coulombs, and $\tau_{\rm s}$ is the time constant of the shaping 
amplifier in seconds. The expression holds for a single stage CR-RC filter. 
A higher order filter (as used by us) reduces the noise by a factor 
of $\sim\,1/1.5$.
We conclude that shot noise associated with the leakage current does not
dominate the 59 keV energy resolution of the detectors.
Although the electronic noise of CZT detectors has been discussed in the 
literature\cite{Nemi:98,Nemi:01,Luke:01,Luke:02}, additional work is required 
to identify the dominant noise source for the special case of pixellated 
Imarad detectors.
\subsection{Photopeak efficiency}
A crucial property of CZT detectors is their photopeak efficiency $\epsilon$. 
We define the latter as the fraction of photoeffect events which end up in the photopeak of the measured energy 
spectrum, i.e.\ within $\pm$ 1 FWHM of the peak of the distribution. Although $\epsilon$ is a crucial 
quantity for all CZT applications, it is often not measured directly. 
The peak to valley ratio is commonly used as a token for $\epsilon$. 
Another common practice is to fit an exponential function to the ``valley'' 
and to extrapolate the fit to lower pulse heights to estimate the fraction 
of counts hidden in the Compton continuum. 
We have used a calibrated Cs$^{137}$ source to determine the absolute 
photopeak efficiency at 662 keV. The source was placed at a sufficiently large distance 
from the detector so that the event rate was low and the dead time was less than 20\%. 
Based on a log-histogram of the time intervals between events, the dead-time has been 
estimated to a fractional accuracy of better than 5\%. 
We determine the photopeak efficiency by comparing the dead-time 
corrected event rate in the photopeak with the theoretical expectation
assuming a perfect detector. Using Method 1 to correct for the DOI dependence 
of the anode signals, we measure a 662~keV photopeak efficiency of 67\%.
The result agrees with earlier estimates at 122 keV\cite{Nari:00}.
We anticipate that the efficiency can be increased by using, in addition 
to the pixel with the largest pulse, the signals from neighboring 
pixels\cite{Du:99,He:00,Nari:02,Kale:02}.
\subsection{Detector simulations}
\begin{figure}
  \centering
  \begin{minipage}{3.5in}
  \includegraphics[width=3.4in]{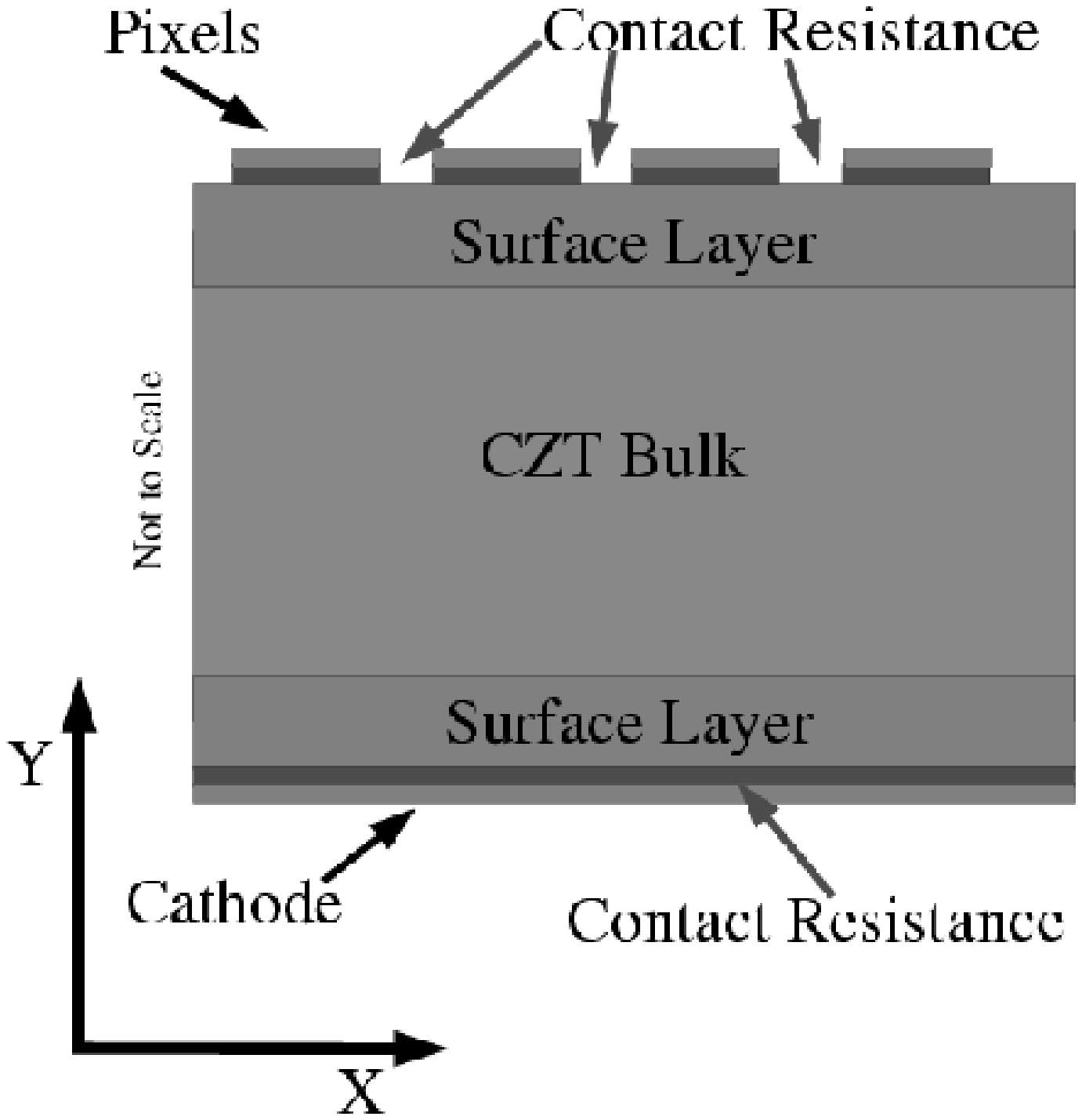}
  \end{minipage}
  \begin{minipage}{2.8in}
  \caption{  \label{model}
  Graphical representation of the simulation model.  
  The simulations account for contact resistance and surface conductivity.
  The vertical dimensions of the contact and surface layers have been 
  enlarged for clarity.}
  \end{minipage}
\end{figure}
Detector simulations can be used to enhance our understanding of the detector
and the measured signals and to optimize the detector design.
We have been developing a simulation chain with 3 components:
\begin{enumerate}
\item We use the GEANT~4 code\cite{Agos:03} developed for high energy 
physics applications to simulate the interaction of the primary
photon and secondary particles with the detector.
For our application, it is important to use GEANT 4 together with the low 
energy extension GLECS/GLEPS which properly accounts for the binding energy of inner electrons.
The GEANT~4 simulations allow us to study in detail the importance and 
properties of Compton events as well as non-local energy loss processes
of secondary electrons. The GEANT 4 simulation outputs the locations and 
magnitudes of energy depositions.
\item We use the commercial semiconductor device simulation package ATLAS\cite{Atlas} 
to determine the electric field and the weighting potentials inside the detector.
The bulk is doped with $1.5\times10^6$ electrons/cm$^{3}$.
We used a surface layer with a higher electron concentration than the 
bulk to simulate enhanced conductivity at the substrate surfaces. 
We also included significant contact resistance based on pixel-cathode and 
pixel-pixel I-V measurements. The contact resistivity of individual pixels 
is on the order of several G$\Omega$.  
The ATLAS package is advertised to handle 3-D simulations.
However, we found this feature modestly useful,
as the simulations converge only if the number of simulated 
grid points and contacts is small.
\item We use our own code to track electrons through the detector and 
to compute the corresponding anode and cathode signals, based on the electric 
field and weighting potentials from the ATLAS simulations.
The code outputs pulses that can be analyzed in the same way as
experimental data.
\end{enumerate}
\begin{figure}
  \centering
\begin{center}
\begin{minipage}{3.5in}
  \includegraphics[width=3.4in]{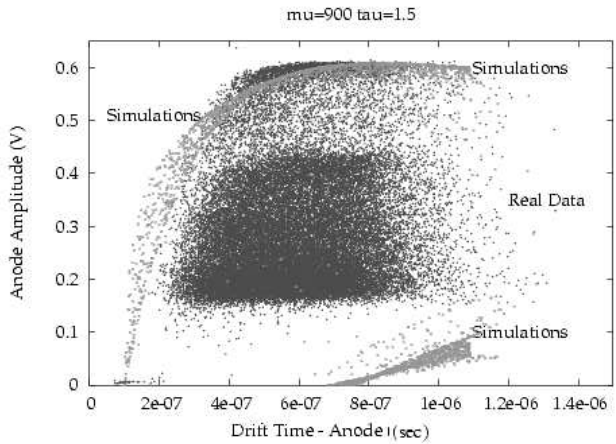}
\end{minipage}
\begin{minipage}{2.8in}
  \caption{\label{xstart}
  Anode amplitude versus the duration of the cathode pulse
  (drift time) for simulated and experimental data. 
  Events at long drift times are simulated well while events at short drift
  times are not.  
  The results indicate the electric field and/or the weighting potential close 
  to the anode pixels are not yet simulated correctly. 
  While there are only photoeffect events in the
  simulations, the real data shows also the Compton continuum.}
\end{minipage}
\end{center}
\end{figure}
Fig.\ \ref{xstart} compares time resolved data (standard Imarad detector, Cs$^{137}$, 662 keV)
to simulated data. In the case of the simulations, only photoeffect events are shown.
The mobility and trapping parameters of the simulations were chosen to
optimize the agreement between the simulations and the data. 
The mobility is given by the longest drift times. The trapping time is
constrained by the slope of the photopeak line at long drift times. 
Based on these simulations we estimate $\mu = 900~\rm cm^2V^{-1}s^{-1}$ and $\tau = 1.9~\mu$sec. 
The results depend on the assumed voltage drop across the contacts, and 
should be considered preliminary.
The simulations deviate from the data in the region of short drift times. 
The results indicate the electric field and/or the weighting potential close 
to the anode pixels are not yet simulated correctly. 
We plan to fine-tune the simulations with a better experimental determination of the
surface conductivity and contact properties, based on a series of 4 point 
measurements, and pixel-pixel and pixel-cathode I-V measurements.
Other effects that need further study are the effects of charge carrier 
diffusion\cite{He:00,Kale:02}, and the contribution of holes to the anode and cathode signals\cite{Nemi:01}. 
\section{Imarad detectors with alternative contacts}
\label{contacts}
Standard Imarad detectors, contacted with an In cathode and In anodes, suffer at low energies 
from electronic noise produced by the detectors. Various authors reported on using blocking cathode and/or 
anode contacts to reduce the noise associated with the leakage current.
Narita et al.\ (1999)\cite{Nari:99} reported good results for 4~mm thick Imarad detectors.
While a standard Imarad detector achieved an energy resolution of 6 keV, 
a detector with Au cathode and CdS anodes and another detector with Au cathode and In anodes 
achieved 59 keV energy resolutions of $\sim$3 keV.
Nemirovski et al.\ (2001)\cite{Nemi:01} tested Imarad detectors with a variety of 
cathode and anode contacts. They achieved the best energy resolution, 5 keV at 122 keV,
with a detector that used a blocking Au cathode and ohmic In anodes. 
A detailed noise analysis indicated that reverse biased Schottky contacts generate 
a similar level of noise as ohmic contacts. However, forward biased Schottky contacts 
generate copious generation-recombination noise and $1/f$ noise.
In a more recent publication, Narita et al.\ (2002)\cite{Nari:02} tested a large number of 
detectors with Au cathodes and In anodes, and with Au cathodes and Au anodes. 
Only 50\% of the detectors showed clear photopeaks. For the good detectors, 
typical 122 keV energy resolutions of $\sim$6 keV were found.

\begin{figure}[t]
  \centering
  \hspace*{-0.9cm}
  \begin{minipage}{3.2in}
  \includegraphics[width=2.8in]{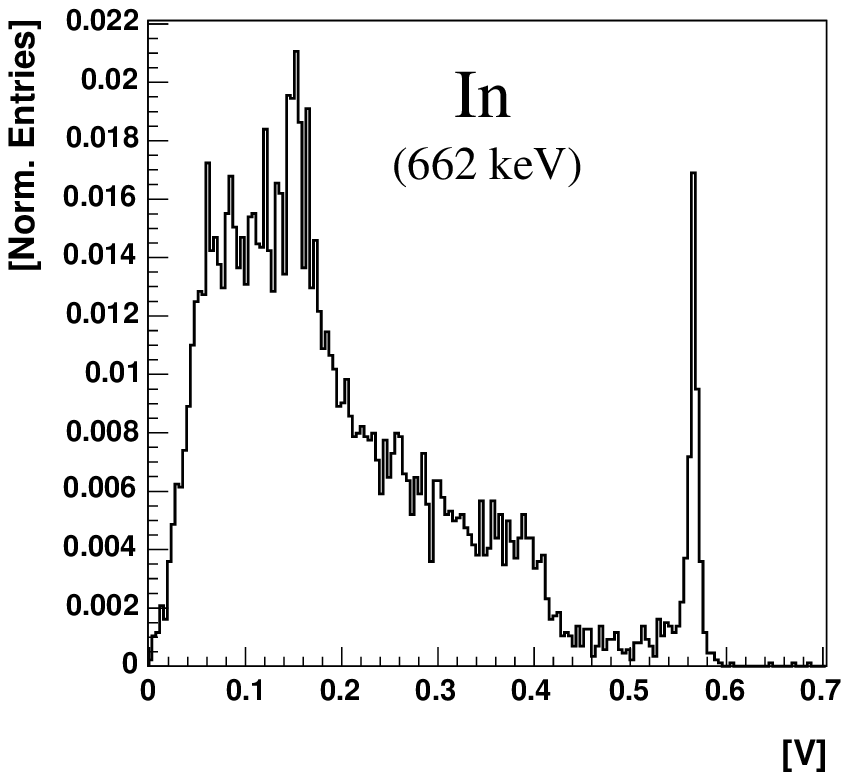}
  \end{minipage}
  \begin{minipage}{3.2in}
  \includegraphics[width=2.8in]{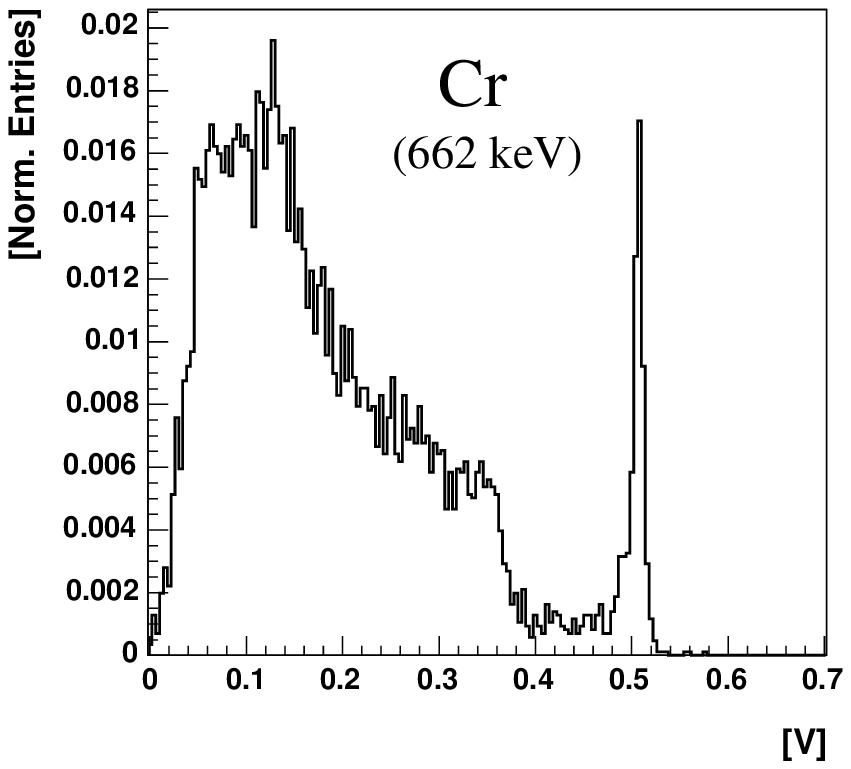}
  \end{minipage}\\[-0.2cm]
  \hspace*{-0.9cm}
  \begin{minipage}{3.2in}
  \includegraphics[width=2.8in]{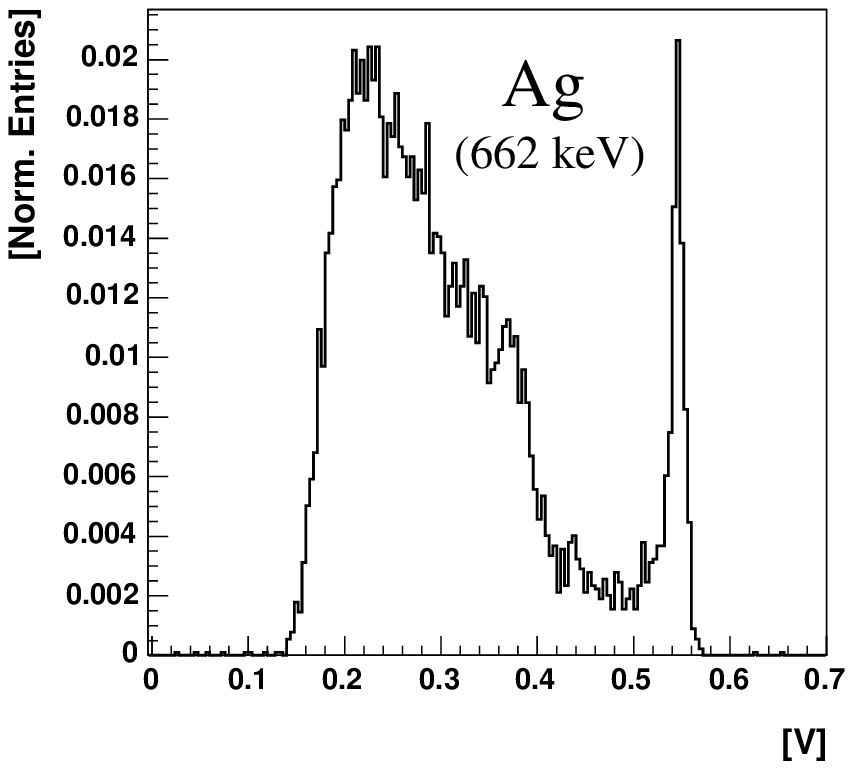}
  \end{minipage}
  \begin{minipage}{3.2in}
  \includegraphics[width=2.8in]{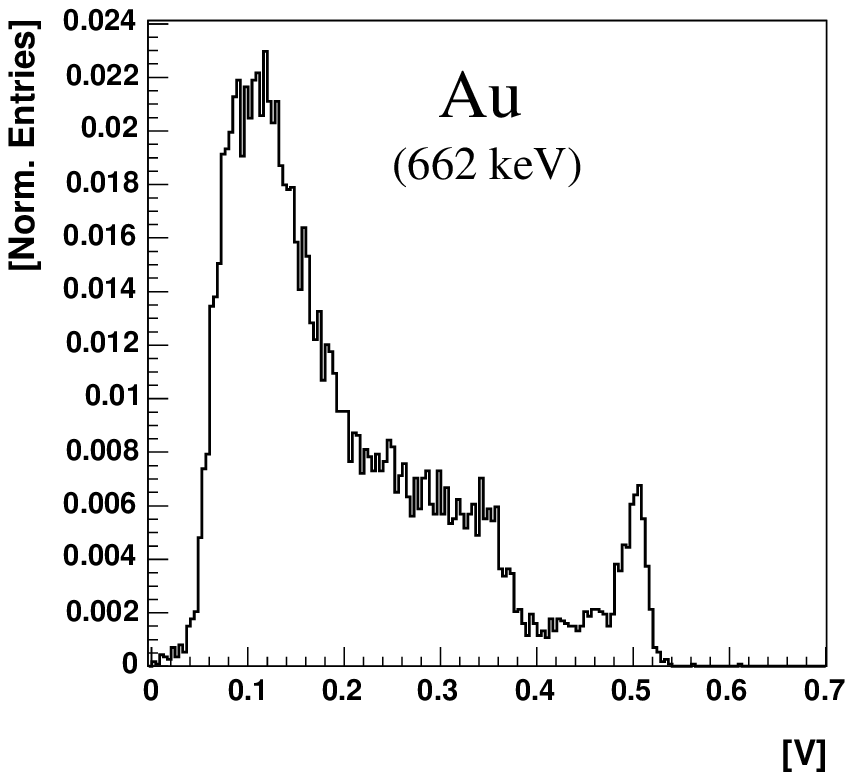}
  \end{minipage}
  \vspace*{-0.7cm}
  \caption{  \label{four}
  662 keV energy spectra of 4 Imarad detectors contacted with In, Cr, Ag, and Au.
  Using Method 1 for the DOI correction, the FWHM energy resolutions are
  10 keV (In), 12 keV (Cr), 17 keV (Ag), and 32 keV (Au). 
  The spectrum of the Ag contacted detector looks a bit different than that of the
  other detectors owing to a higher trigger threshold used during the measurements.
  Bias voltages of $\sim$-1000 V were used.}
\end{figure}
In the following, we present results from contacting CZT detectors with different metals. 
Here we have chosen the high workfunction metals Cr, Ag, Au, and Pt that are expected to make 
reverse biased Schottky contacts on the cathodes of the n-type Imarad CZT. 
In the case of the Cr, Ag, and Au contacted detectors, we used the same
metal for the anode and the cathode contacts. 
A fourth detector, called Pt--In detector in the following, was fabricated from a standard
Imarad detector by polishing off the In cathode and replacing it with a Pt cathode, while
leaving the In anode contacts untouched. 

Using a custom-machined 1.6 mm square punch and 0.3~mm brass foil, we fabricated several masks with 8$\times$8 
pixels of 1.6~mm pixel diameter and 2.4~mm pixel pitch.
The Cr and Ag contacts were applied at Washington University.
After polishing the detectors with 0.05 $\mu$m particle size alumina 
suspension, the detectors were etched for 2 min in 2\% Bromide, 20\% Lactic Acid, 
dissolved in Ethylene Glycol\cite{Wrig:02}. After etching, the samples were rinsed in Methanol. 
Metallization was performed with an electron beam evaporator. 
The Au and Pt contacts were applied at Fisk University. 
The Au detector used a blank Imarad detector, with the surfaces polished and prepared by Imarad.
The substrate was submerged for 30 sec in 1\% Bromide in Methanol, and subsequently rinsed in pure Methanol. 
The Pt--In detector was fabricated from a standard Imarad detector.
The cathode was polished off with 0.05 $\mu$m particle size alumina suspension.
After polishing, the cathode side was etched 2 min in 1\% Bromide in Methanol
and subsequently rinsed in Methanol. The In contacts were not modified.
The contacts were deposited with a sputter system.

Energy spectra (662 keV) from relatively good pixels of the 
standard Imarad detector (In), and the Cr, Ag, and Au contacted detectors are shown in Fig.\ \ref{four}. 
The four detectors show well pronounced photopeaks, and the mean signal amplitudes are very similar
for the four detectors. The Pt-In contacted detector showed a rather poor performance, 
and is not included in the figure.
At 662 keV, the In contacted detector performs best with an energy resolution of 10 keV. 
The Cr and Ag contacted detectors exhibit resolutions of 12 keV and 16 keV, respectively. 
The Au and Pt--In contacted detectors give substantially poorer high energy resolutions of 
32 keV and 92 keV, respectively. 
At 59 keV, we find energy resolutions of 7 keV (In), 11 keV (Cr), 17 keV (Ag), and 5 keV (Au).
The Pt contacted detector does not exhibit a clear 59 keV photopeak.
The results are summarized in Table \ref{contact}.
\begin{table}[t]
\begin{center}
{\small 
\begin{tabular} {ccccc}\hline
Cathode       & Anode    & Contact    &{ FWHM(662 keV)} & { FWHM(59 keV)} \\ 
Material            & Material & Deposition & & \\ \hline 
In & In & Imarad    & 10 keV   &  7 keV   \\ 
Cr & Cr & E-Beam    & 12 keV   & 11 keV   \\ 
Ag & Ag & E-Beam    & 17 keV   & 17 keV   \\ 
Au & Au & Sputtered & 32 keV   & 5 keV    \\ 
Pt & In & Sputtered & 92 keV   & --       \\ \hline
\end{tabular}
}
\caption{FWHM energy resolutions of Imarad detectors contacted with different metals. 
The Pt--In contacted detector did not show a clear 59 keV photopeak.}
\label{contact}
\end{center}
\end{table}
For each detector we have tested several pixels. The standard Imarad detector and the 
Cr and Ag contacted detectors exhibit consistent behavior with 
a $\sim$10\% spread in energy resolutions.
The Pt and Au contacted detectors show a larger spread, with some 
pixels not giving any photopeaks either at 662 keV or at 59 keV, 
or at both energies.

The Ag, Au, and Pt--In detectors exhibit low leakage currents around 1 nA that
are approximately constant from 200 V to 1500 V. The Cr detector shows a low 
leakage current of $\sim$3~nA when the cathode is positively biased with respect to the anode. 
We measure a large leakage current of 30 nA/1000 V at negative cathode bias.
The detector should be refabricated with flipped detector orientation.
\section{Summary and outlook}
\label{summary}
CZT has emerged as the detector material of choice for the detection of 10 keV 
to 600 keV X-rays without the need for cryogenic cooling. 
CZT is having a major impact on the field of hard X-ray astronomy.
In this paper we have presented measurements of Imarad CZT standard detectors.
We have shown that the cathode amplitude is better suited to correct 
the anode amplitudes for the DOI than the duration of the cathode pulse.
After correction for the DOI, the detectors achieve a photopeak efficiency 
of 67\% at 662 keV. The 59 keV and 662 keV energy resolutions 
are 7 keV and 10 keV, respectively.
We have described first results from contacting Imarad detectors with high workfunction
metals. Although the detectors exhibit low leakage currents, they perform
in most cases worse than standard Imarad detectors.
Together with our finding that the standard Imarad detectors are not limited by
shot noise from the leakage current, our study indicates that the main emphasis of the
detector optimization should be on finding contacts that do not generate electronic
noise themselves, rather than focusing on contacts that minimize the leakage current.

\begin{figure}[tb]
\begin{center}
\begin{minipage}[htb]{7.5cm}
{\includegraphics[width=6.4cm]{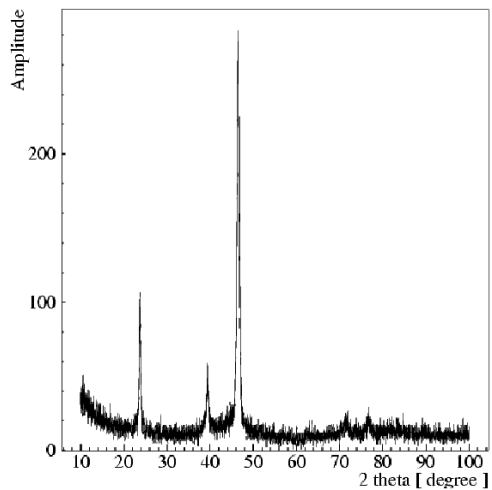}}
\end{minipage}
\begin{minipage}[htb]{7.5cm}
\caption{\label{xd} The panel shows the X-ray $\theta$-2$\theta$ diffraction scan of the 
``anode'' face of an Imarad CZT detector  taken at Washington University. 
The diffraction peaks show that the anode and cathode surfaces of the Imarad detectors are
oriented parallel to the (111) plane.}
\end{minipage}
\end{center}
\end{figure}

In a collaboration of Washington University, Fisk University, and UCSD 
(J.~Matteson, R.~T.~Skelton), we have started a systematic study of CZT 
surface preparation, contact materials, contact deposition, annealing, 
and passivation techniques. 
The aim is to improve on the low-energy performance 
of the detectors and to achieve a high yield of good detectors.
\begin{figure}[bt]
\begin{center}
\includegraphics[width=14cm]{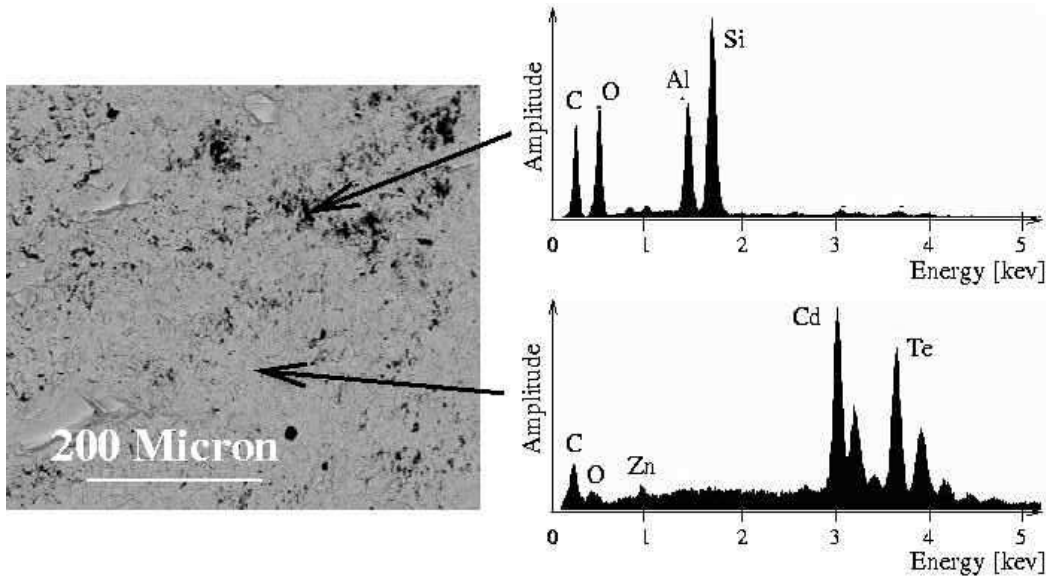}
\end{center}
\vspace*{-0.8cm}
\caption{
\label{sem} The left panel shows a secondary electron microscope (SEM) image
of an Imarad CZT detector taken at Washington University. 
Light regions show ``heavy'' elemental composition, while dark regions show
``lighter'' elemental composition. The right hand panels show two energy spectra
of backscattered electrons taken from small (several microns diameter) regions
with widely different chemical composition. 
Contamination of the CZT surface with Al and Si polish residuals can be recognized.}
\end{figure}
To steer our developments into the right direction, we are guiding the optimization 
process with several diagnostic tools:\\[1.5ex]
{\it Secondary Electron Microscope (SEM):} SEM allows us to map the chemical composition of the
CZT surfaces. The energy spectrum of backscattered electrons from the light regions 
makes it possible to determine the relative Ca, Zi and Te compositions 
to an accuracy of 1\%-2\%. As an example, Fig.\ \ref{sem} shows a SEM picture of an Imarad detector 
after polishing; dark (light) regions correspond to a light (heavy) composition. 
The analysis of $\sim$1 micron diameter dark regions 
shows that the substrate is heavily contaminated with Al and Si residuals from the 
polishing process, and should be cleaned more thoroughly.\\[1.5ex]
{\it X-ray diffraction:} We have been examining the CZT with X-ray diffraction for crystal 
structure and orientation. In the future, defect-induced broadening of the $\theta$-$2\theta$ peaks 
will also be studied. An example of an X-ray $\theta$-$2\theta$ diffraction scan from an Imarad
CZT crystal is shown in Fig.\ \ref{xd}.
The line structure shows that the crystal is oriented with the (111) direction
perpendicular to the anode and cathode side of the detector.
The diffraction peaks do not show the K-$\alpha$1 and K-$\alpha$2 doublet from the Cu source 
proving that line broadening owing to crystal irregularities is significant.\\[1.5ex]
{\it Photoluminescence, photoconductivity, and Pockels effect:} 
Photoluminescence measurements use a laser beam tuned above the band gap energy that 
is directed onto the CZT surface and the emitted radiation is recorded as a function of 
wavelength below the band gap energy. Radiative recombination from the band 
edge states as well as defect and impurity levels can be resolved. 
The technique is very sensitive to defect structure and impurities from the 
intrinsic material as well as from any subsequent surface preparation\cite{Leop:86}.
Photoconductivity measurements can be used to measure the surface recombination 
rate\cite{Cui:02}. These measurements can be combined with internal field 
mapping using the Pockels effect\cite{DeAn:96} and the response to alpha particles.  
We aim to decouple the charge loss contributions of the bulk ($\mu-\tau$ product) 
and the surface (owing to fabrication related recombination effects).  
\section*{Acknowledgment}
We thank Uri El Hanany from Imarad for several free CZT detectors. HK is grateful to J.~Matteson 
for very detailed discussion of many key characteristics of CZT detectors.
It is a pleasure to acknowledge Randy Korotev for the cross calibration of 
radio active sources, and Christine Floss for help with the SEM measurements.
Thanks to electrical engineer P.~Dowkontt, and electrical technician 
G.~Simburger for their work. 
We acknowledge L.~Sobotka, D.~Leopold, and J.~Buckley for helpful discussions. 
This work has been supported by NASA under contracts NNG04WC176 and 
NNG04GD70G, and the NSF/HRD grant no.\ 0420516 (CREST).


\begin{thebibliography}{1}
\bibitem{Giac:62}{Giacconi, R., Gursky, H., Paolini, F.~R., et al.\  1962, Phys.\ Rev.\ Lett., 9, 439}  
\bibitem{Giac:79}{Giacconi,R., Branduardi, G., Briel, U., et al.\ 1979, ApJ., 230, 540}
\bibitem{True:83}{Tr\"umper, J.\ 1983, Adv.\ Space.\ Res., 2, No.\ 4, 241}
\bibitem{Garm:03}{Garmire, G.\ P., Bautz, M.\ W., Ford, P.\ G., Nousek, J.\ A., Ricker, G.\ R.\ Jr.\ 
2003, Proc.\ SPIE, 4851, 28}
\bibitem{Stru:01}{Str\"uder, L., Briel, U., Dennerl, K., et al.\ 2001, A\&A, 265, L18}
\bibitem{Matt:78}{Matteson, J.~L.\ 1978, Proc.\ AIAA 16th Aerospace Sciences Meeting, 78} 
\bibitem{Brad:90}{Bradt, H.~V., Swank, J.~H., Rothshild, R.~E.\ 1990, Adv.\ Space Res., 10, 297}
\bibitem{Piro:95}{Piro, L., Scarsi, L., Butler, R.~C.\ 1995, SPIE Proc., 2517, 169}
\bibitem{Uber:03}{Ubertini, P., Lebrun, F., Di Cocco, G., et al.\ 2003, A\&A, 411, L131}
\bibitem{Gehr:00}{Gehrels, N.\ 2000, Proc.\ SPIE, 4140, 42}
\bibitem{BeyondEinstein}{NASA's Structure and Evolution of the Universe 2003 Roadmap, 
Beyond Einstein: From the Big Bang to Black Holes, http://universe.gsfc.nasa.gov/}
\bibitem{conX}{The Constellation Project Team, ``Constellation-X Technology Readiness and Implementation Plan (TRIP)
Report'', http://constellation.gsfc.nasa.gov/images/science/resources/documents/TRIPReport\_NoCost.pdf}
\bibitem{Sala:04}{Salamon, M.~H., private communication}
\bibitem{Grin:03}{Grindlay, J.E.,  Craig, W.W., Gehrels, N., et al.\ 2003, Proc.\ SPIE, 4851, 331}
\bibitem{McCo}{McConnell, M., Ryan, J., Macri, J.\ 2004, Talk at the ``Beyond Einstein:
From the Big Bang to Black Holes'' meeting, Stanford Linear Accelerator Center,
Stanford University, 12-15 May 2004}
\bibitem{eV}{eV Products, 373 Saxonburg Boulevard, Saxonburg, PA 16056}
\bibitem{Imarad}{Imarad Imaging Systems, Ltd., Rabin Park, 10 Plaut Street, Rehovot, Israel}
\bibitem{Bicr}{Bicron, 6341 San Ignacia Road, San Jose, CA 95119}
\bibitem{Taka:02}{Takahashi, T., Mitani, T., Kobayashi, Y., et al.\ 2002, IEEE Trans.\ Nucl.\ Sci., 49, 1297}
\bibitem{Bolotnikov01}{Bolotnikov, A.E., Chen, C.M.H., Cook, W.R., et al.\ 2002, IEEE Trans.~Nucl.~Sci., 49, 1941}
\bibitem{Macr:02}{Macri, J.~R., Donmez, B., Hamel, L.~A., et al.\ 2003, 
IEEE Nucl.~Sci.~Symposium Conf.~Record, 02CH37399, 468}
\bibitem{Nari:99} Narita, T., Bloser, P., Grindlay, J., Jenkins, J., and Yao, H. 1999, Proc.\ SPIE, 3768, 55
\bibitem{Perk:03}{Perkins, J.~S., Krawczynski, H., Dowkontt, P., 
Proc.\ of the 13th International Workshop on Room-Temperature Semiconductor X- and Gamma-Ray Detectors, 
IEEE Meeting, Portland (OR), 2003, astro-ph/0310910}
\bibitem{Nari:00}{Narita, T., Bloser, P., Grindlay, J., Jenkins, J.A.\ 2000, Proc.\ SPIE 4141, 89}
\bibitem{Varn:96}{Varnell, L.~S., Mahoney, W.~A., Hull, E.~L., Butler, J.~F.\ 1996, Proc.\ SPIE, 2806, 424}
\bibitem{Wong:96}{Wong, A.~S., Harrison, F. A., Varnell, L.~S.\ 1996, Proc.\ SPIE, 2806, 442}
\bibitem{Hull:97}{Hull, E., Pehl, R., Varnell, L.\ 1997, IEEE Trans. Nucl. Sci., 44, 870}
\bibitem{Fran:99}{Franks, L.~A., Brunett, A.~B., Olsen, R.~W., et al.\ 1999, NIMA, 428, 95}
\bibitem{Slav:00}{Slavis, K.~R., Dowkontt, P., Duttweiler, F., et al.\ 2000, Proc.\ SPIE, 4140, 249}
\bibitem{Pars:96}{Parsons, A., Barthelmy, S., Bartlett, L.\ 1996, Proc.\ SPIE, 2806, 432}  
\bibitem{Jenk:03}{Jenkins, J.\ A., Tomohiko, N., Grinday, J.~E.\ 2003, Proc.\ SPIE, 4851, 866}
\bibitem{Mura:03}{Murakami, M.~M., Kobayashi, Y., Kokubun, M., et al.\ 2003, IEEE Trans.\ Nucl. Sci., 50, 1013}
\bibitem{Frab:03}{Fraboni, B., Cavallini, A., Dusi, W.\ 2003, IEEE 0-7803-8258-7/03}
\bibitem{Arms:99a}{Armstrong, W. T., Colborn, L. B., Dietz, L. K., Ramsey, D. B.\ 1999, 
Astrophys.\ Letters and Communications, 39, 413 }
\bibitem{Arms:99}{Armstrong, T.~W., Colborn, B.~L., Ramsey, B.~D.\ 1999, Report SAIC-TN, 990115R3}
\bibitem{Perf:01}{Perfect, C.~L., Bird, A.~J., Dean, A.~J., et al.\ 2001, Astrophys.\ \& Spa.\ Sci., 276, 263} 
\bibitem{Blos:02}{Bloser, P.F., Narita, T., Jenkins J.A., et al.\ 2002, Proc.\ SPIE, 4497, 88}
\bibitem{Li}Li, W., He, Z., Knoll, G.F., et al.\ 2001, NIMA, 458, 518
\bibitem{Nemi:01}{Nemirovski, Y., Asa, G., Gorelik, J., Peyser, A.\ 2001, NIMA, 458, 325} 
\bibitem{Hong:02} Hong, J., Bellm, E.~C., Grindlay, J.~E.,Narita, T.\ 2003, Proc.\ SPIE, 5165, 54
\bibitem{Barret} Barret, H.H., Eskin, J.D., Barber, ,H.B. 1995, Phys. Rev. Lett., 75, 156
\bibitem{Luke} Luke, P.N. 1995, ,In: Procs. of the ``9th International Workshop on
Room Temperature Semiconductor X- and Gamma-Ray Detectors, Associated
Electronics and Applications'', Grenoble,  France,  18-22 Sept., 1995
\bibitem{Nemi:98}{Nemirovski, Y., Asa, G., Ruzin, A., Gorelik, J., Sudharsanan, R.\ 1998, 
Journ.\ Electr.\ Materials, 27, 6} 
\bibitem{Luke:01}{Luke,P.N., Amman, M., Lee, J.S., Manfredi, P.F.\ 2001, 
IEEE Trans.\ Nucl.\ Sci., 48, 282}
\bibitem{Luke:02}{Luke, P.N., Lee, J.S., Amman, M., Yu, K.M.\ 2002,
IEEE Trans.\ Nucl.\ Sci., 49, 1950}
\bibitem{Du:99}{Du, Y.\ F., et al.\ 1999, IEEE Trans.\ Nucl. Sci., NS-46, 844}
\bibitem{He:00}{He, Z., Li, W., Knoll, G.~F., Wehe, D.~K., Du, Y.~F.\ 2000, NIMA, 439, 619}
\bibitem{Nari:02}{Narita, T., Grindlay, J.~E., Jenkins J. A., et al.\ 2002, Proc.\ SPIE, 4497, 79}
\bibitem{Kale:02}{Kalemci, E., Matteson, J.\ L.\ 2002, NIMA, 478, 527} 
\bibitem{Agos:03}{Agostinelli, S., Allison, J., Amako, K., et al.\ 2003, 
NIMA, 506, 250}
\bibitem{Atlas} SILVACO International, Inc., 4701 Patrick Henry Drive, Building 2,
Santa Clara, CA 95054
\bibitem{Wrig:02}{Wright, G., Cui, Y., Roy, U.~N., et al.\ 2002, IEEE Trans.~Nucl.~Sci., 49, 2521}
\bibitem{Leop:86}{Leopold, D.~J., Ballingall, J.~M., Wroge, M.~L., 1986,
Appl.\ Phys.\ Lett., 49, 1473}
\bibitem{Cui:02}{Cui, Y., Groza, M., Hillman, D., Burger, A., James, R.~B.\ 2002, 
J.~Appl.~Phys., 92, 2557-2560}
\bibitem{DeAn:96}{De Antonis, et al.\ 1996, IEEE Trans.~ Nucl.~ Sci., 43, 1487}
\end{thebibliography}
\end{document}